\documentstyle[PASJadd]{PASJ95}
\input epsf.sty

\begin{document}
\title{Near-infrared emission-line galaxies in the Hubble Deep Field North}

\author{
Fumihide {\sc Iwamuro},
Kentaro {\sc Motohara},
Toshinori {\sc Maihara},
Jun'ichi {\sc Iwai},\\
Hirohisa {\sc Tanabe},
Tomoyuki {\sc Taguchi},
Ryuji {\sc Hata},
Hiroshi {\sc Terada},
and
Miwa {\sc Goto}\\
{\it Department of Physics, Kyoto University, Kitashirakawa, Kyoto 606-8502}\\
{\it E-mail(FI): iwamuro@cr.scphys.kyoto-u.ac.jp}\\
\\ [6pt]
Shin {\sc Oya}\\
{\it Communications Research Laboratory, Koganei, Tokyo 184-8975}\\
\\ [6pt]
Masanori {\sc Iye},
Michitoshi {\sc Yoshida},
and
Hiroshi {\sc Karoji}\\
{\it Optical and Infrared Astronomy Division, National Astronomical Observatory, Mitaka, Tokyo 181-8588}\\
and\\
Ryusuke {\sc Ogasawara},
and
Kazuhiro {\sc Sekiguchi}\\
{\it Subaru Telescope, National Astronomical Observatory, 650 North Aohoku Place, Hilo, HI 96720, USA}\\
}
\abst{
We present the 2.12~$\mu$m narrow-band image of the Hubble Deep Field North
taken with the near-infrared camera (CISCO) on the Subaru telescope. 
Among five targets whose H$\alpha$ or [O~{\sc iii}] emission lines are redshifted 
into our narrow-band range expected from their spectroscopic redshift,
four of them have strong emission lines, especially for the two [O~{\sc iii}]
emission-line objects. The remaining one target shows no H$\alpha$
emission in spite of its bright rest-UV luminosity, indicating that 
this object is already under the post-starburst phase. The volume-averaged
$SFR$ derived from the detected two H$\alpha$ emission is roughly consistent
with that evaluated from the rest-UV continuum.}

\kword{cosmology: observations --- galaxies: formation --- infrared: galaxies}

\maketitle
\thispagestyle{headings}

\section
{Introduction}
The Hubble Deep Field North (HDF-N) has been the deepest and the most scrutinizingly
observed area since the first images taken by WFPC2 (Williams et al. 1996).
The acquired data dramatically pushed the study of the global star-formation
history of the early Universe with the establishment of the technique to
determine the photometric redshift. The volume-averaged star-formation rate
($SFR$) derived from UV-to-optical data in this field shows a declination
from the peak at $z\simeq$ 1.5 toward the higher-$z$ (Madau et al. 1996),
while there are some suggestions that the $SFR$ remains almost at a constant 
value when the reddening correction (Calzetti 1998; Steidel et al. 1999) 
or the selection effect of the rest-UV surface brightness
(Pascarelle et al. 1998) is taken into account. On the other hand, the
volume-averaged $SFR$ derived from the H$\alpha$ luminosity density tends to
be larger than that evaluated from the rest-UV continuum (Pettini et al. 1998; 
Glazebrook et al. 1999; Yan et al. 1999), possibly due to a difference
in the reddening effect between the UV continuum and the H$\alpha$ emission
line.

Now, the HDF-N has become a standard area for studying distant galaxies, 
and various catalogs are available (Fern\'andez-Soto et al. 1999; Thompson 
et al. 1999; etc.). In this paper, we report on the results of the observation 
of the HDF-N by the near-infrared camera CISCO mounted 
on the Subaru telescope. This observation was planned not only for 
verifying the system performance in the engineering phase just after
the first light observing run, but also for studying the H$\alpha$ or
[O~{\sc iii}] luminosity density in this field. The details of the observation 
and the data reduction are described in section 2, the detection of the 
near-infrared emission-line objects and the numerical results are 
reported in section 3, and the properties of these objects and the volume-averaged 
star-formation rate are discussed in section 4. Throughout this paper, we 
assume the Hubble constant of $H_0=50$ km s$^{-1}$ Mpc$^{-1}$ and the 
deceleration parameter of $q_0=0.5$.\\

\section
{Observations and Data Reduction}
Observations of the HDF-N were made at the Subaru telescope on 1999 February 23,25,
and 27 with the Cooled Infrared Spectrograph and Camera for OHS (CISCO; Motohara 
et al. 1998). The field was imaged through a standard $K'$ (1.96--2.30~$\mu$m)
and H$_2$ 1--0 S(1) (2.110--2.130~$\mu$m, hereafter {\it N212}) filters. Within the 
$2'\times 2'$ field-of-view of the CISCO, there are five candidates whose rest-frame 
optical emission lines are redshifted into the bandpass of the {\it N212} filter on the 
basis of their spectroscopic redshifts.  and several more candidates expected similarly 
from their photometric redshifts (Fern\'andez-Soto et al. 1999 and references therein).
The images were taken using an octagonal-star-like dither pattern with a diameter 
of 12$''$. In the $K'$-band case, 20~s$\times$12 sequential exposures were made 
at each position (60~s$\times$6 in the case of {\it N212}), yielding a total exposure
time of 1920~s for the standard round of the observation. Unfortunately, the 
seeing size was unstable during this observing run, and varied from 0$.\hspace{-2pt}''$4 
to 0$.\hspace{-2pt}''$9 in each set of exposures. The observations are summarized 
in table~1.


All image reductions were carried out using IRAF. The initial reduction to make a
``quick look image'' followed that of standard infrared-image processing: a flat 
fielding using the ``standard $K'$ sky flat'', median-sky subtraction, and 
shift-and-add to combine the image. Then, the positions of the bright sources were
listed in a ``mask table'' and the median-sky image was reconstructed by masking 
out the listed bright objects. In the $K'$-band case, the median smoothed sky image 
was also used as the self-sky-flat image, and all images were re-flattened by this frame.
After applying secondary self-sky flat fielding and sky subtraction, the following 
reduction processes were carried out: a bad-pixel correction, self-fitting to remove any 
minor residual of the bias pattern with masking the listed bright object, 
and shift-and-add of all images with a 3~$\sigma$ clipping algorithm.
The reduced $K'$ and {\it N212} images are shown in figure~1. The effective exposure times
are equal to the total exposure time of 5280~s in the $K'$- and 9360~s in the 
{\it N212}-band, respectively, because there was no discarded frame. The FWHMs for the point source 
in the final images are 0$.\hspace{-2pt}''$58 ($K'$) and 0$.\hspace{-2pt}''$67 ({\it N212}).


\section
{Results}
After a Gaussian smoothing was applied to the $K'$ image to make the image size identical 
to that of the {\it N212} image, object detection was performed on each frame using 
SExtractor (Bertin, Arnouts 1996). We employed the ``BEST'' (Kron-like) magnitude
output as the total magnitude, while the {\it N212}$-K'$ color within 1$.\hspace{-2pt}''$16~$\phi$ (10~pix) 
aperture was measured by running SExtractor in the {\it double-image 
mode}: $K'$ magnitudes for the {\it N212} selected sample were derived using the 
{\it N212} frame as a reference image for detection and the $K'$ image for measurements 
only and {\it vice versa} for the $K'$ selected sample. 
Next, each catalog was compared with the redshift catalog of Fern\'andez-Soto et al. (1999)
under the condition that the nearest corresponding pair having the position difference  
smaller than 0$.\hspace{-2pt}''$9 is regarded as the same object. The size of the effective
area is 3 arcmin$^2$ and the distribution of the identified objects is shown in figure~2.


The identified and unidentified objects are plotted on the {\it N212}$-K'$ color versus {\it N212} 
magnitude diagram (figure~3) together with the simulated data of the artificial point 
sources whose intrinsic {\it N212}$-K'$ colors are 0 mag. Here, the horizontal {\it N212} mag is 
the aperture magnitude, because the scatter of the simulated data becomes smaller than 
in the case when the total magnitude is applied. The 98\% confidence level estimated from 
the simulated data points is expressed as
\begin{equation} {\it N212}-K'=2.5~{\rm log}(1\pm 10^{{\it N212}-21.45 \over 2.5}), \end{equation}

\noindent
corresponding to the 2.3 $\sigma$ detection limit of the emission line of 2.6$\times 10^{-20}
\rm ~W~m^{-2}$. 

Figure~4 shows the redshift distribution of the identified objects. Among the objects with
available spectroscopic redshift, seven are located where the emission lines are 
redshifted into the {\it N212} filter bandpass. The SEDs of three objects (A,B,C) located 
in the redshift range for H$\alpha$ are shown in figure~5a. The brightness and 
the shape of these SEDs are quite similar, while the H$\alpha$ strength of object C 
is far different from that of A or B. The SEDs of other two objects (D,E) for [O~{\sc iii}]
are also shown in figure~5b. We could not detect any emission from the remaining two objects 
for redshifted H$\beta$ and the other emission-line candidates expected from the photometric 
redshifts. The intensity of the emission lines and the $SFR$s for objects A--E are listed in table~2. 
Here, we assumed an emission line ratio of H$\alpha$/[N~{\sc ii}]~6583,6548=2.3 (Kennicutt 1992; 
Gallego et al. 1997; Yan et al. 1999), and employed the conversion factor from Kennicutt (1998): 
$SFR_{\rm H\alpha}$($M_{\odot}$~yr$^{-1})=7.9\times 10^{-42}L$(H$\alpha$) (erg~s$^{-1}$),
$SFR_{\rm UV}$($M_{\odot}$~yr$^{-1})=1.4\times 10^{-28}L_{\nu}$(1600~\AA) (erg~s$^{-1}$Hz$^{-1}$).


\section
{Discussion}
Recently, Meurer et al. (1999) reported that the UV spectral index $\beta$ ($f_\lambda 
\propto \lambda^\beta$) is a good indicator of dust extinction. They found 
a simple relationship between the dust absorption at 1600~\AA ($A_{1600}$) and the 
$\beta$-parameter: $A_{1600}=4.43+1.99\beta$. We assume that this correlation is also applicable 
to our results. First, we checked our results by using the same diagram of Meurer et al. 
(1999) (figure~6). The vertical axis is the ``normalized line flux'' defined as the ratio 
of line flux to flux density $f_{1600}$. In this diagram, the ratio of H$\alpha /f_{1600}$ 
indicates the rough duration time of post-starburst 
phase, because the H$\alpha$ decays with a shorter timescale ($<$ 20~Myr) than that 
for the UV continuum ($<$ 100~Myr) (Kennicutt 1998). The condition of 
$SFR_{\rm H\alpha}$ = $SFR_{\rm UV}$ without dust extinction implies a constant $SFR$ of 
the galaxy, displayed as the solid line in figure~6a. We employed the 'net' obscuration
expression of the stellar continuum reported by Calzetti (1998) to correct for the reddening 
factor at the wavelength of H$\alpha$. As can be seen, objects A and B
are probably undergoing a starburst phase with a constant $SFR$ in contrast to object C,
which has already finished starburst activity. In the case of the [O~{\sc iii}] emission line 
(figure~6b), objects D and E are located in a higher region compared with the distribution 
of the local starburst galaxies, and reside in a region similar to other Lyman-break galaxies. 
Although the [O~{\sc iii}] emission line is not a good quantitative $SFR$ indicator, due to its 
large diversity (Kennicutt 1992), it is still notable that the star-forming galaxies at 
high-$z$ tend to have strong [O~{\sc iii}] emission lines. 

The remaining data points correspond
to the other candidates from the photometric redshifts (open symbols) and the two redshifted 
H$\beta$ (small filled symbols) converted to the H$\alpha$ flux assuming the case B condition.
Although one marginal value with 1$<S/N<$2.3 is included (open circle in figure~6a), the 
accuracy of the photometric redshift ($\sim$~10\%) does not seem to be sufficient to select the emission-line 
candidates for a narrow-band survey like this. On the other hand, one of two data points 
corresponding to the redshifted H$\beta$ emission is located at a higher position than 
the constant $SFR$ line in figure~6a. This is also a marginal data point, however, indicating
that this object is a possible young starburst galaxy.


Next, the volume-averaged $SFR$s were calculated using the H$\alpha$ luminosity function
at $z~\sim$~0.7--1.9 obtained by Yan et al. (1999). The 2.3 $\sigma$ detection limit of 
2.6$\times 10^{-20}\rm ~W~m^{-2}$ corresponds to the H$\alpha$ luminosity of $L_{\rm lim}=
6.4\times 10^{41}\rm ~erg~s^{-1}$. The survey volume ($V$) is $585\ \rm Mpc^3$ for the
redshifted H$\alpha$ emission line. The sum of the $SFR_{\rm H\alpha}$ for the 
two H$\alpha$-detected objects ($SFR_{\rm sum}$) is 22.5 $M_{\odot}$~yr$^{-1}$ (see table~2).
We then calculated the volume-averaged $SFR$ ($\rho^\star$) using the following formulae:
\begin{eqnarray}
\alpha&=&-1.35,\\
L^\star&=&7\times 10^{42}\ \rm erg\ s^{-1},\\
\phi^\star&=&1.7\times 10^{-3}\ \rm Mpc^{-3},\\
L_{\rm sum}(x)&=&\int_x^\infty L\phi^\star \left(\frac{L}{L^\star}\right)^\alpha {\rm exp}\left(\frac{L}{L^\star}\right)d\left(\frac{L}{L^\star}\right),\\
\rho^\star&=&\left(\frac{{SFR}_{\rm sum}}{V}\right)\left(\frac{L_{\rm sum}(0)}{L_{\rm sum}(L_{\rm lim})}\right)\nonumber\\
&=&0.05~M_{\odot}\rm ~yr^{-1}Mpc^{-3}.
\end{eqnarray}

Our value is slightly larger than the previous results derived from 
the rest-UV continuum at $z~\sim$~2.2 (see figure~7), showing the same tendency
as the previous H$\alpha$ survey reported by Glazebrook et al. (1999) or Yan et al. (1999). 
If we take the reddening effect expected from the $\beta$-parameter into account, 
the resultant value of 0.09 $M_{\odot}$~yr$^{-1}$Mpc$^3$ is almost consistent with the 
reddening-corrected value calculated from the rest-UV continuum (Calzetti 1998; 
Steidel et al. 1999). Although the error is relatively large, the reddening correction
by the $\beta$-parameter works well as far as the present results are concerned.

Finally, we calculated the [O~{\sc iii}] luminosity density in the same manner as mentioned above 
to compare with the H$\alpha$ luminosity density expected from the previous results for 
the UV continuum. Because the resultant ratio of the [O~{\sc iii}]/H$\alpha$ luminosity density is
about 5, it cannot be explained as being the same star-forming activity in the local universe.
The most conceivable cause is that the high-$z$ star-forming galaxies emit a larger amount 
of ionizing photons than local starburst galaxies due to some different type of 
activity, making the [O~{\sc iii}] lines brighter. There is also another possibility,
that oxygen becomes the major cooling species, owing to the deficiency of metal elements,
similar to the nearby blue-compact-dwarf galaxies with metallicity of $\sim$~0.1 solar 
(Kunth, Ostlin 1999). Infrared spectroscopy is necessary to investigate such activity in 
detail, and further observations by the OH-airglow Suppressor spectrograph on the Subaru 
telescope (Maihara et al. 1993; Iwamuro et al. 1994) are expected to provide the infrared 
spectra of high-$z$ star forming galaxies.\\


\section
{Conclusions}
The 2.12~$\mu$m narrow-band image of the HDF-N was taken with the near-infrared 
camera on the Subaru telescope. The main conclusions of the present paper are the 
following:
\begin{enumerate}
\item We examined three objects whose H$\alpha$ lines should be redshifted into our {\it N212} 
narrow-band filter as expected from their known spectroscopic redshifts of $z~\sim$~2.2. We found 
that two of the three targets are undergoing a starburst phase similar to the local 
starburst galaxies, while the last one target has already finished starburst 
activity, since no H$\alpha$ emission was detected from this object.

\item We also detected strong [O~{\sc iii}] emission lines from both targets with the 
known spectroscopic redshift of $z~\sim$~3.2. We infer that the high-$z$ star-forming galaxy 
tends to show strong [O~{\sc iii}] emission, just like the Lyman-break galaxies reported 
by Pettini et. al (1998).

\item The volume-averaged $SFR$ derived from the detected H$\alpha$ lines is slightly
larger than the values previously reported from the rest-UV continuum. They become almost 
consistent when the reddening correction is applied using the UV spectral index $\beta$.
\end{enumerate}

\vspace{2pc}\par

The present result was accomplished during an engineering observing run of the Subaru 
Telescope, and is indebted to all members of Subaru Observatory, NAOJ, Japan.  
We would like to express thanks to the engineering staff of Mitsubishi Electric Co. 
for their fine operation of the telescope, and the staff of Fujitsu Co. for the timely 
provision of control software. 

\section*{References}

\re Bertin E., Arnouts S. 1996, A\&AS 117, 393
\re Calzetti D. 1998, in Proc. the Ringberg Workshop on Ultraluminous Galaxies (astro-ph/9902107)
\re Connolly A.J., Szalay A.S., Dickinson M., Subbarao M.U., Brunner R.J. 1997, ApJ 486, L11
\re Fern\'andez-Soto A., Lanzetta K.M., Yahil A. 1999, ApJ 513, 34
\re Gallego J., Zamorano J., Arag\'on-Salamanca A., Rego M. 1995, ApJ 455, L1
\re Gallego J., Zamorano J., Rego M., Vitores A.G. 1997, ApJ 475, 502
\re Glazebrook K., Blake C., Economou F., Lilly S., Colless M. 1999, MNRAS 306, 843
\re Iwamuro F., Maihara T., Oya S., Tsukamoto H., Hall D.N.B., Cowie L.L., Tokunaga A.T., Pickles A.J. 1994, PASJ 46, 515
\re Kennicutt R.C.Jr. 1992, ApJ 388, 310
\re Kennicutt R.C.Jr. 1998, ARA\&A 36, 189
\re Kunth D., Ostlin G. 2000, A\&AR in press (astro-ph/9911094)
\re Lilly S.J., Le Fevre O., Hammer F., Crampton D. 1996, ApJ 460, L1
\re Madau P., Pozzetti L., Dickinson M.E. 1998, ApJ 498, 106
\re Madau P., Ferguson H.C., Dickinson M.E., Giavalisco M., Steidel C.C., Fruchter A. 1996, MNRAS 283, 1388
\re Maihara T., Iwamuro F., Hall D.N.B., Cowie L.L., Tokunaga A.T., Pickles A.J. 1993, Proc. SPIE 1946, 581
\re Meurer G.R., Heckman T.M., Calzetti D. 1999, ApJ 521, 64
\re Motohara K., Maihara T., Iwamuro F., Oya S., Imanishi M., Terada H., Goto M., Iwai J. et al. 1998, Proc. SPIE 3354, 659
\re Pascarelle S.M., Lanzetta,K.M., Fern\'andez-Soto A. 1998, ApJ 508, L1
\re Pettini M., Kellogg M., Steidel C.C., Dickinson M., Adelberger K.L., Giavalisco M. 1998, ApJ 508, 539
\re Steidel C.C., Adelberger K.L., Giavalisco M., Dickinson M., Pettini M. 1999, ApJ 519, 1
\re Teplitz H.I., Malkan M.A., McLean I.S. 1999, ApJ 514, 33
\re Thompson R.I., Storrie-Lombardi L.J., Weymann R.J., Rieke M.J., Schneider G., Stobie E., Lytle D. 1999, AJ 117, 17
\re Tresse L., Maddox S.J. 1998, ApJ 495, 691
\re Treyer M.A., Ellis R.S., Milliard B., Donas J., Bridges T.J. 1998, MNRAS 300, 303
\re Williams R.E., Blacker B., Dickinson M., Dixon W.V.D., Ferguson H.C., Fruchter A.S., Giavalisco M., Gilliland R.L., et al. 1996, AJ 112, 1335
\re Yan L., McCarthy P.J., Freudling W., Teplitz H.I., Malumuth E.M., Weymann R.J., Malkan M.A. 1999, ApJ 519, L47

\onecolumn
\begin{table*}[t]
\begin{center}
Table~1.\hspace{4pt}{Observation log.}\\
\end{center}
\vspace{6pt}
\begin{tabular*}{\textwidth}{@{\hspace{\tabcolsep}
\extracolsep{\fill}}p{8pc}ccc}
\hline\hline\\ [-6pt]
Date&Filter&Exposure&Seeing\\
[4pt]\hline\\[-6pt]
1999 February 23&$K'$      &1680~s(20~s$\times$12$\times$7)&0$.\hspace{-2pt}''$5--0$.\hspace{-2pt}''$8\\
                &{\it N212}&4320~s(60~s$\times$6$\times$12)&0$.\hspace{-2pt}''$5--0$.\hspace{-2pt}''$9\\
[4pt]\hline\\[-6pt]
1999 February 25&$K'$      &1920~s(20~s$\times$12$\times$8)&0$.\hspace{-2pt}''$5--0$.\hspace{-2pt}''$9\\
                &{\it N212}&5040~s(60~s$\times$6$\times$14)&0$.\hspace{-2pt}''$6--1$.\hspace{-2pt}''$0\\
[4pt]\hline\\[-6pt]
1999 February 27&$K'$      &1680~s(20~s$\times$12$\times$7)&0$.\hspace{-2pt}''$4--0$.\hspace{-2pt}''$8\\
[4pt]\hline\\[-6pt]
\end{tabular*}
\end{table*}

\begin{table*}[t]
\begin{center}
Table~2.\hspace{4pt}{Emission-line strength and $SFR$.}\\
\end{center}
\vspace{6pt}
\begin{tabular*}{\textwidth}{@{\hspace{\tabcolsep}
\extracolsep{\fill}}p{6pc}ccccc}
\hline\hline\\ [-6pt]
Object&Spectroscopic&Line&$F$(line)&$SFR_{\rm H\alpha}$&$SFR_{\rm UV}$\\
&redshift&&10$^{-20}$W~m$^{-2}$&\multicolumn{2}{c}{$M_{\odot}$~yr$^{-1}$}\\
[4pt]\hline\\[-6pt]
A&2.237&H$\alpha$+[N~{\sc ii}]&5.0$\pm$1.1 &9.8$\pm$2.2   &5.2\\
B&2.233&H$\alpha$+[N~{\sc ii}]&6.5$\pm$1.1 &12.7$\pm$2.2  &7.3\\
C&2.232&H$\alpha$+[N~{\sc ii}]&$<$1.1      &$<$2.2        &0.7\\
D&3.233&[O~{\sc iii}]         &5.7$\pm$1.1 &(69.3$\pm$13.7)$^\ast$&5.9\\
E&3.216&[O~{\sc iii}]         &10.0$\pm$1.1&(121.7$\pm$13.7)$^\ast$&18.5\\
[4pt]\hline\\[-6pt]
\end{tabular*}
\vspace{6pt}\par\noindent
$*$ $SFR_{\rm [O~III]}$($M_{\odot}$~yr$^{-1})=1.5\times 10^{-41}L$([O~{\sc iii}]) (erg~s$
^{-1}$) is assumed (Teplitz et al. 1999).
\end{table*}

\begin{figure}[p]
\epsfxsize=11cm

\epsfxsize=11cm
\caption{$K'$(top) and {\it N212}(bottom) images of the HDF-N. The candidates whose rest-frame 
optical emission lines are redshifted into the bandpass of the {\it N212} filter are indicated
by circles labeled A--E.}
\end{figure}
\begin{figure}[p]
\epsfxsize=17cm
\epsfbox{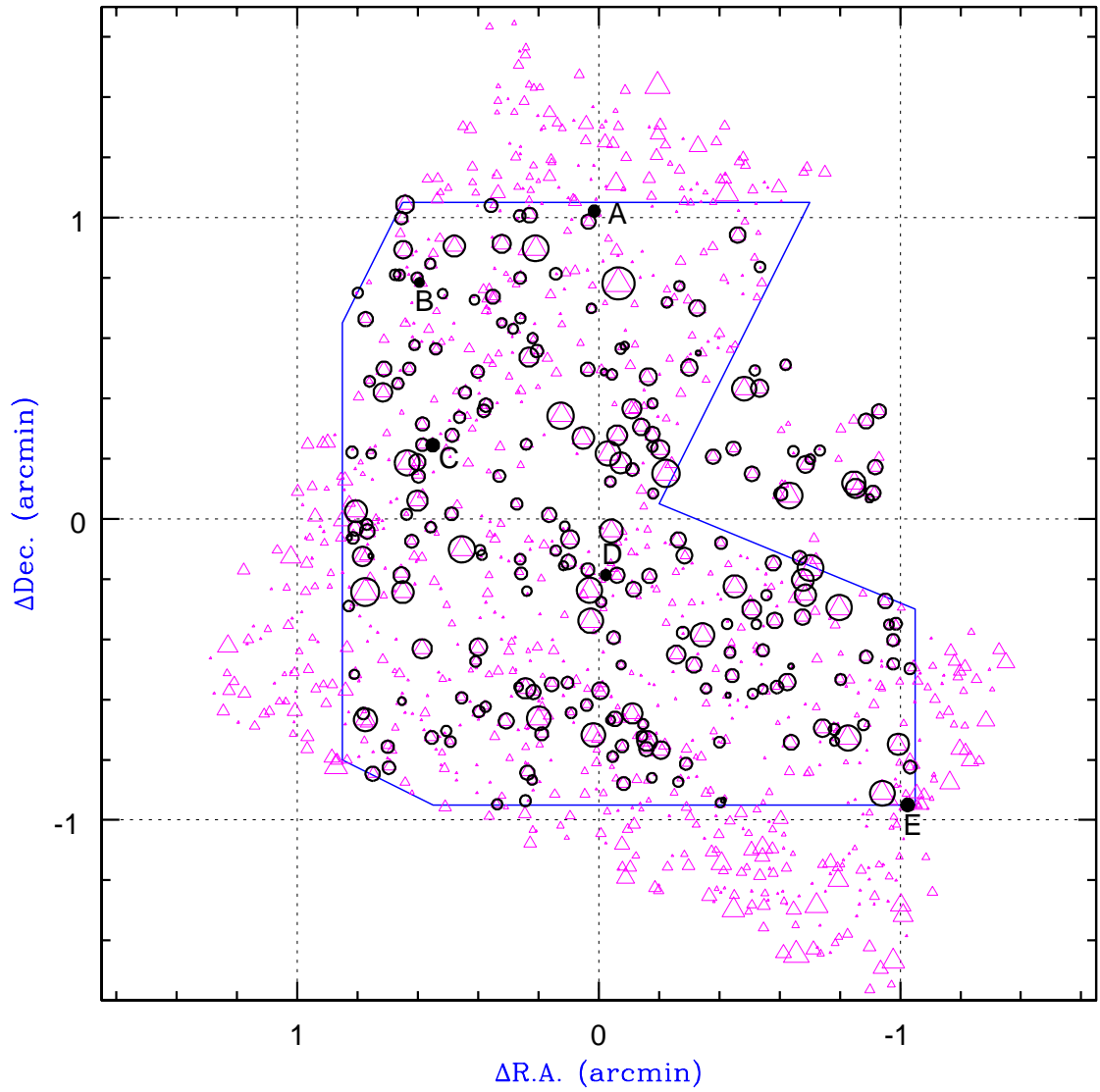}
\caption{Distribution of the identified objects (thick circle) using the redshift catalog of 
Fern\'andez-Soto et al. (1999) (thin triangle). The filled circles denote our five targets
in figure~1. The size of the symbols corresponds to the $K'$ magnitudes.}
\end{figure}
\begin{figure}[p]
\epsfxsize=17cm
\epsfbox{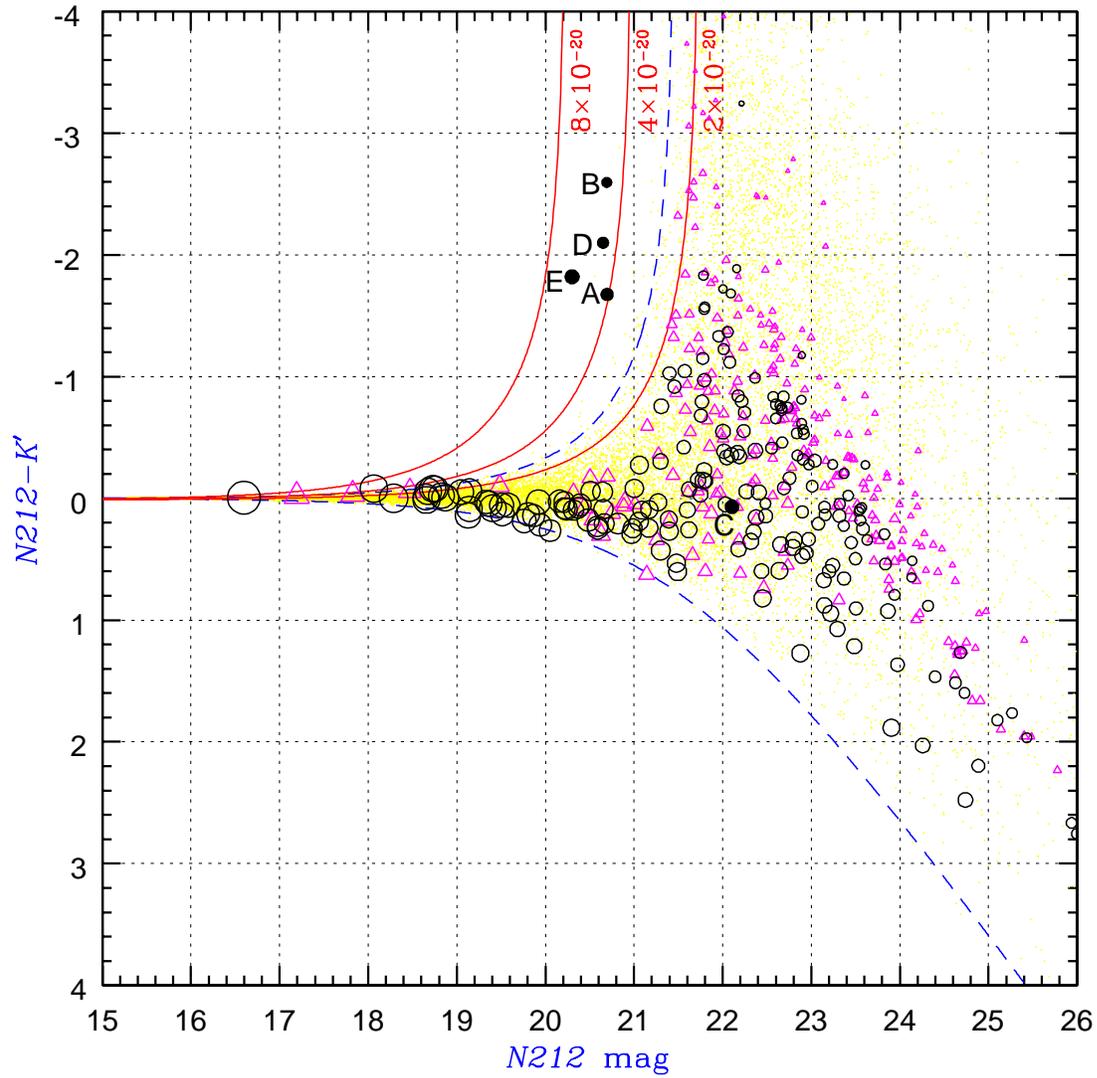}
\caption{{\it N212}$-K'$ aperture color versus {\it N212} magnitude diagram for all the object detected in either frame.
The circles are the identified objects shown in figure~2, while the triangles are unidentified objects.
The filled circles denote our five targets in figure~1. The size of the symbols corresponds to the $K'$ magnitudes.
The simulated data points are represented by thin dots, and the 98\% confidence level is indicated
by the broken lines. The solid lines show the flux of emission line within 1$.\hspace{-2pt}''$16~$\phi$ 
aperture with the unit of W~m$^{-2}$.}
\end{figure}
\begin{figure}[p]
\epsfxsize=17cm
\epsfbox{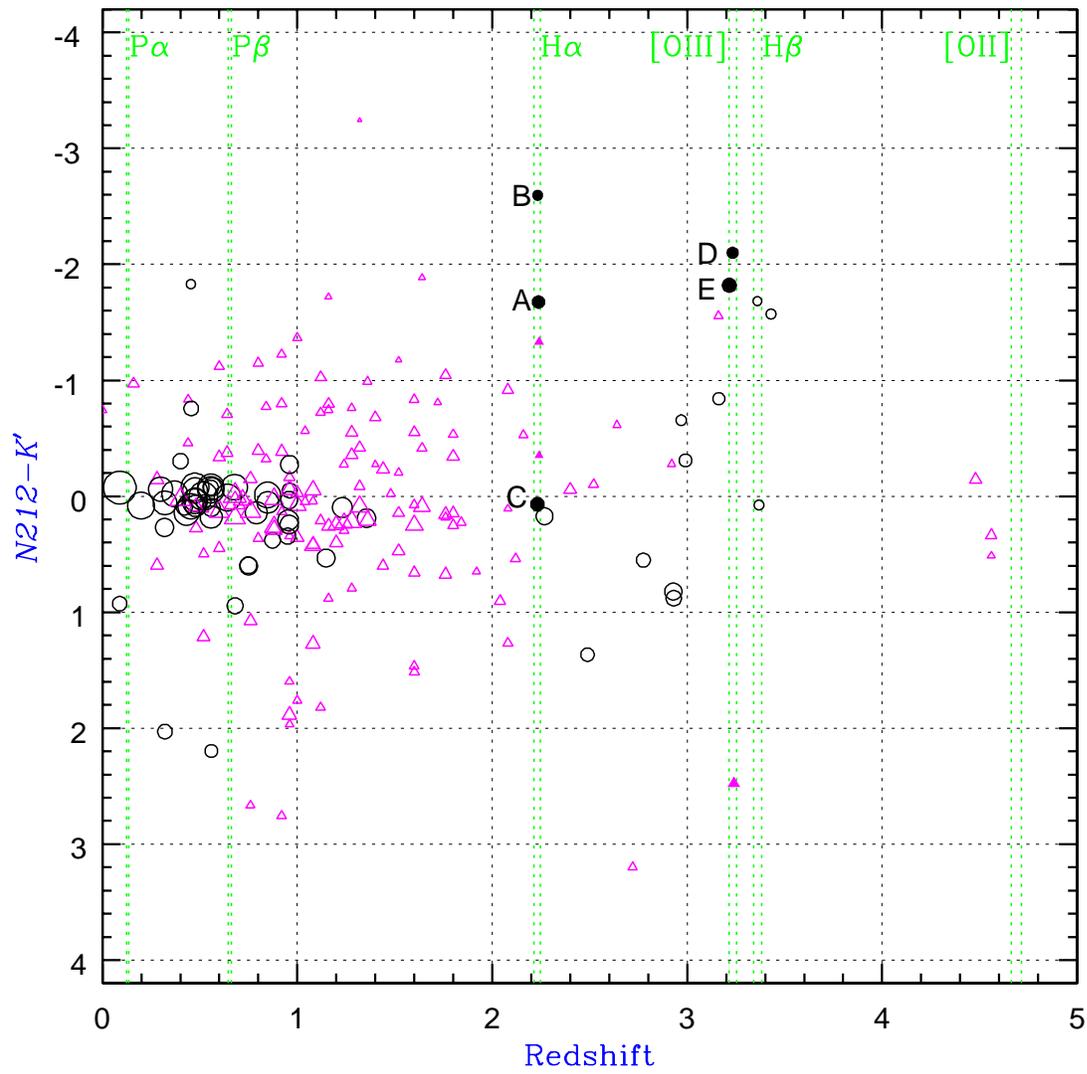}
\caption{Redshift distribution of the identified objects. The circles are plotted based on the spectroscopic redshifts, 
while the triangles are the photometric redshifts calculated by Fern\'andez-Soto et al. (1999) for the 
data whose spectroscopic redshifts are not available. The filled circles denote our five targets
in figure~1. The size of the symbols corresponds to the $K'$ magnitudes.}
\end{figure}
\begin{figure}[p]
\epsfxsize=12cm
\hspace*{2.5cm}\epsfbox{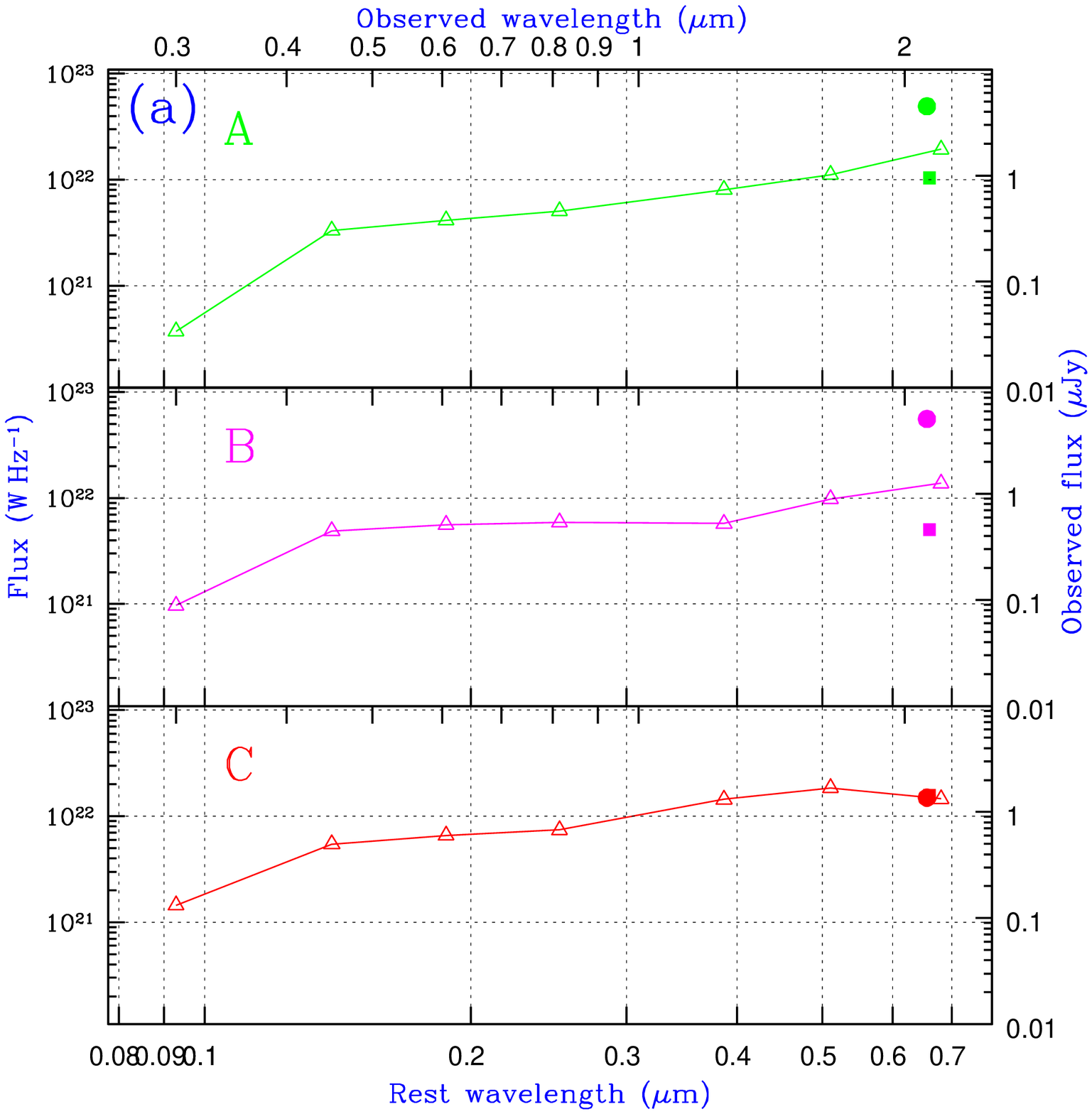}
\vspace{-3cm}

\epsfxsize=12cm
\hspace*{2.5cm}\epsfbox{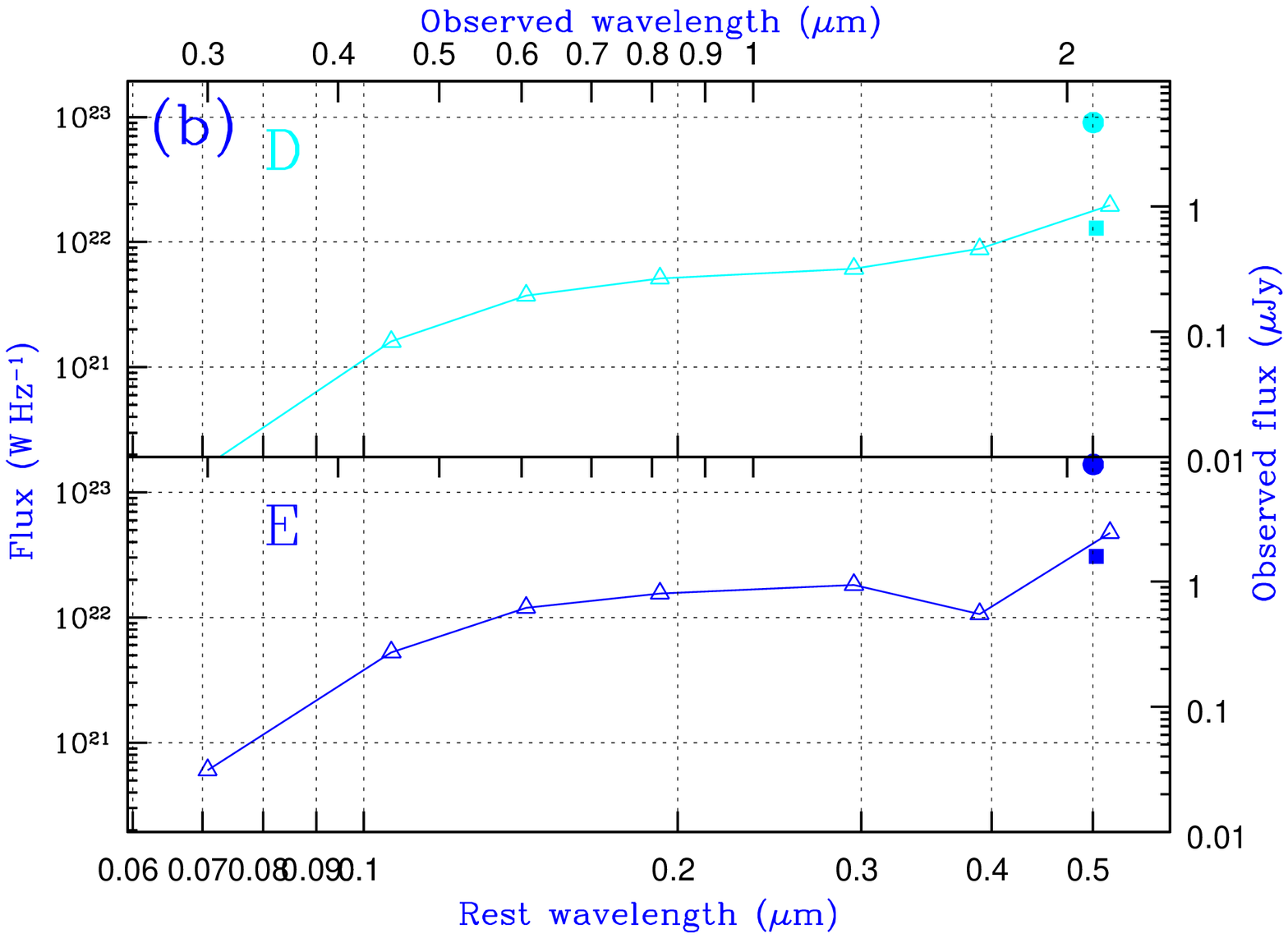}
\caption{(a) SEDs of the objects A--C whose H$\alpha$ lines expected to be redshifted into the 
bandpass of the {\it N212} filter. The triangles are the photometric data by Fern\'andez-Soto et al. 
(1999), and the filled circle and filled square represent of our photometric results for 
{\it N212}- and $K'$-band respectively. (b) The SEDs of the objects D and E with the [O~{\sc iii}]
emission lines. Symbols are the same in these figures.}
\end{figure}
\begin{figure}[p]
\epsfxsize=17cm
\epsfbox{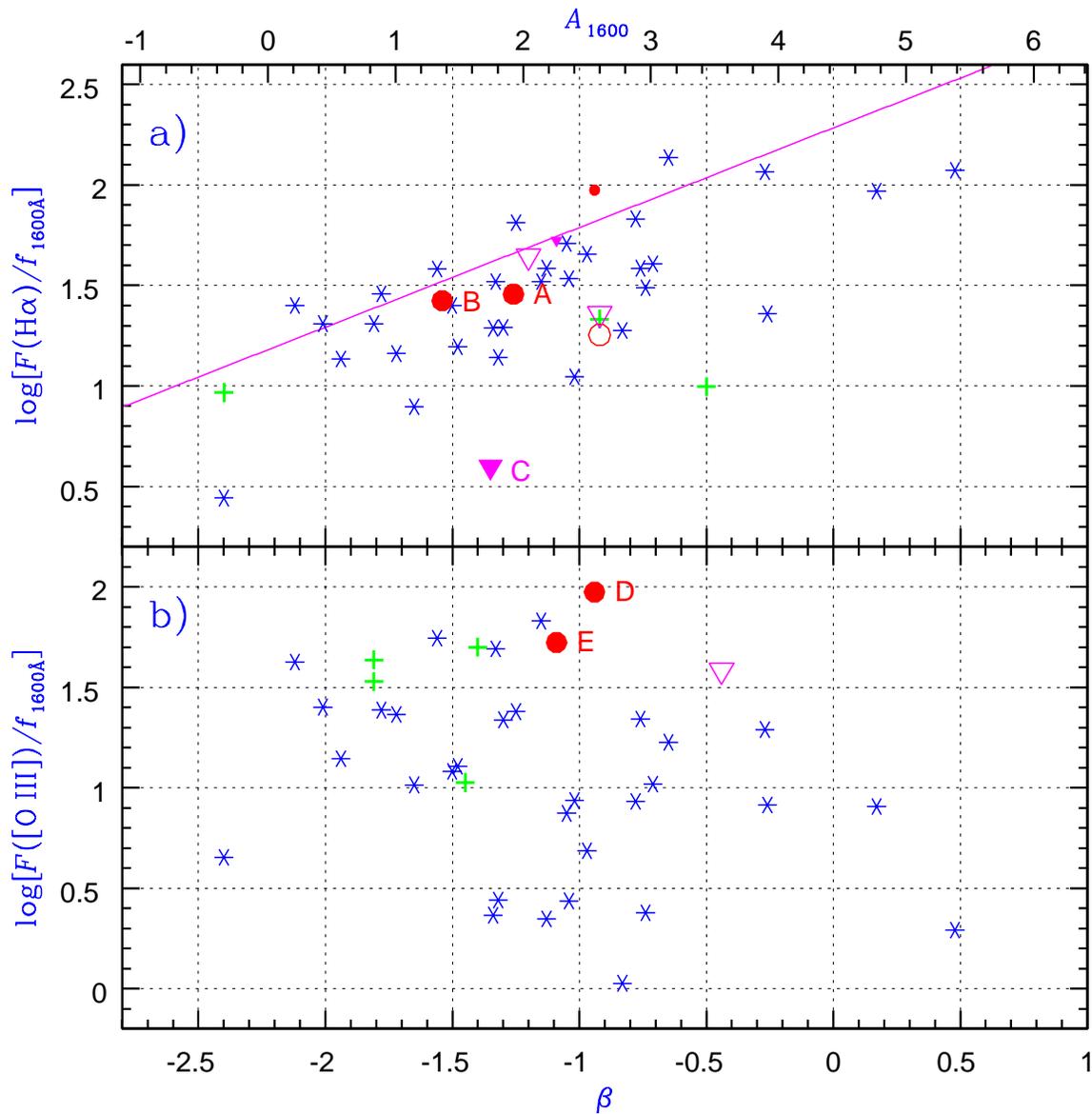}
\caption{Line flux normalized by UV flux density compared to UV spectral index $\beta$. 
The asterisks show a local sample of UV selected starbursts and the thick cross represent 
seven Lyman-break galaxies (Meurer et al. 1999 and references therein). The filled symbols labeled A--E are
the five targets in figure~1, while open symbols are other candidates whose H$\alpha$ or [O~{\sc iii}]
lines expected to be redshifted into the bandpass of the {\it N212} filter from their photometric 
redshift data calculated by Fern\'andez-Soto et al. (1999). The circles indicate the emission-line
data detected with more than a 1~$\sigma$ confidence level, while the inverted triangle correspond to
the 1~$\sigma$ upper limit for the data in which the targeted emission line was not detected.
The small symbols are the two redshifted H$\beta$ data points (see figure~4) converted to the 
H$\alpha$ flux by assuming the case-B condition.}
\end{figure}
\begin{figure}[p]
\epsfxsize=17cm
\epsfbox{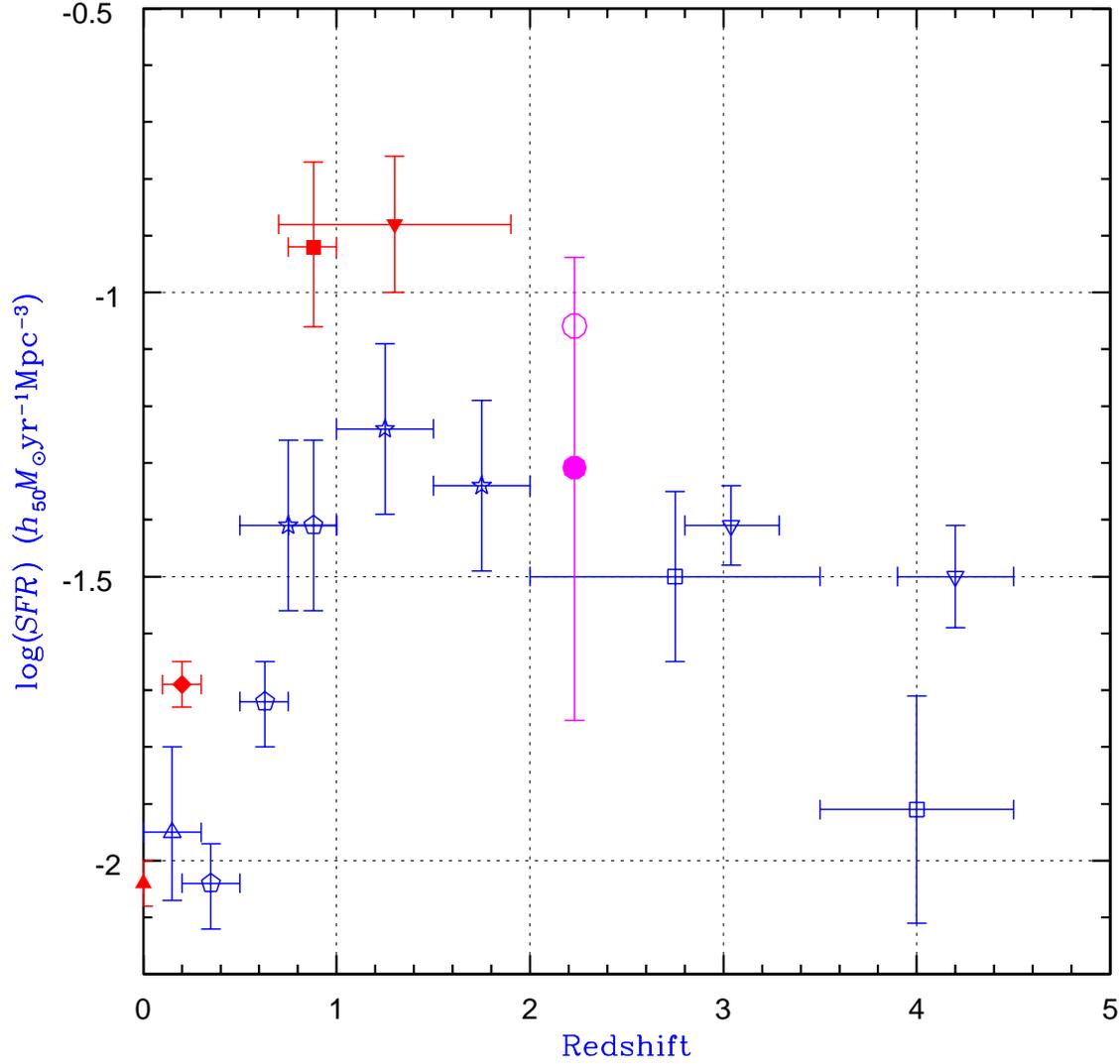}
\caption{Volume-averaged $SFR$ derived by various observing techniques. The points by the rest-UV continuum 
are Treyer et al. (1998; open triangle); Lilly et al. (1996; open pentagons); Connolly et al. (1997; open stars); 
Madau et al. (1998; open squares); Steidel et al. (1999; inverted open triangles). 
The points by the rest-optical emission lines are Gallego et al. (1995; filled triangle); Tresse,
Maddox (1998; filled diamond); Glazebrook et al. (1999; filled square); Yan et al. (1999; inverted
filled triangle); this work (filled circle); after a reddening correction using $\beta$-parameter 
(open circle).}
\end{figure}
\end{document}